\documentclass[12pt,preprint]{aastex}
\usepackage{emulateapj5}

\begin{document}

\def\ast{\mathchar"2203} \mathcode`*="002A 
\def\larrow{\leftarrow} \def\rarrow{\rightarrow} 
\def\=={{\equiv}} 
\def\vinf{v_{\infty}} 
\def\solar{\odot} 
\def\Mdot{\dot M} 
\def\mdot{\dot M} 
\def\zp{{z^{\prime}}} 
\def\rp{{r^{\prime}}} 
\def\Lbol{{L_{bol}}} 
\def\Ro{{R_{o}}} 
\def\Rstar{{\relax \ifmmode {R_\ast} \else $R_\ast$\fi}} 
\def\rstar{{\relax \ifmmode {R_\ast} \else $R_\ast$\fi}} 
\def\rosat{{\it ROSAT\,}}
\def\bbxrt{{\it BBXRT}}
\def \asca{{\it ASCA}}
\def \axaf{{\it AXAF}}
\def \zpup{\hbox{$\zeta$ Pup}}
\def \fx{\relax \ifmmode {f_{\rm X}} \else $f_{\rm X}$\fi}
\def \Lx{\relax \ifmmode {L_{\rm X}} \else $L_{\rm X}$\fi}
\def \lx{\relax \ifmmode {L_{\rm X}} \else $L_{\rm X}$\fi}
\def \lxlb{\relax \ifmmode {L_{\rm X}/L_{\rm Bol}} \else $L_{\rm
    X}/L_{\rm Bol}$\fi} 
\def \Lbol{\relax\ifmmode{L_{\rm Bol}}\else $L_{\rm Bol}$\fi}
\def \lb{\relax\ifmmode{L_{\rm Bol}}\else $L_{\rm Bol}$\fi}

\def\Msunyr{\hbox{${\rm M}_\odot\,$  yr $^{-1}$}}
\def\msunyr{\hbox{${\rm M}_\odot\,$  yr $^{-1}$}}
\def \taunu{\relax \ifmmode {\tau_{\nu}} \else $\tau_{\nu}$ \fi}
\def \sigmanu{\relax \ifmmode {\sigma_{\nu}} \else $\sigma_{\nu}$ \fi}
\def \ecma{\hbox{$\epsilon$ CMa}}
\def \einstein{{\it Einstein}}
\def \pspc{{\rm PSPC}}
\def \euve{{\it EUVE}}
\def \chandra{{\it Chandra}}
\def \xmm{{\it XMM}}
\def \cf{{\it c.f.}}
\def \eg{{\it e.g.}}
\def \etal{et~al.}
\def \ie{{\it i.e.}}
\def \mdovervinf{\relax \ifmmode {\dot M}/v_{\infty} \else ${\dot M}/v_{\infty}$\fi}
\def\Exp{{e}} 
\def\ltwig{\mathrel{\spose{\lower 3pt\hbox{$\mathchar"218$}} 
     \raise 2.0pt\hbox{$\mathchar"13C$}}} 
\def\gtwig{\mathrel{\spose{\lower 3pt\hbox{$\mathchar"218$}}  
     \raise 2.0pt\hbox{$\mathchar"13E$}}} 
\def\spose#1{\hbox to 0pt{#1\hss}} 
\def\blankline{\par\vskip \baselineskip}

\title{X-Ray Line Profiles from Parameterized Emission within an Accelerating Stellar Wind}

\author{Stanley P. Owocki}
\affil{Bartol Research Institute, University of Delaware,
  Newark, DE 19716 \\ E-mail:  owocki@bartol.udel.edu}
\vspace{0.1in}
\author{David H. Cohen}
\affil{Department of Physics and Astronomy, Swarthmore College, Swarthmore, PA 19081 \\
 E-mail:  dcohen1@swarthmore.edu}

\begin{abstract}

Motivated by recent detections by the \xmm\/ and \chandra\/ satellites
of X-ray line emission from hot, luminous stars, we present synthetic
line profiles for X-rays emitted within parameterized models of a
hot-star wind.
The X-ray line emission is taken to occur at a sharply defined
co-moving-frame resonance wavelength, which is Doppler-shifted by a
stellar wind outflow parameterized by a `beta' velocity law,
$v(r)=\vinf (1-\Rstar/r)^\beta$.
Above some initial onset radius $R_o$ for X-ray emission, the radial
variation of the emission filling factor is assumed to decline as a
power-law in radius, $f(r) \sim r^{-q}$.
The computed emission profiles also account for continuum absorption
within the wind, with the overall strength characterized by a
cumulative optical depth $\tau_\ast$.
In terms of a wavelength shift from line-center scaled in units of the
wind terminal speed $\vinf$, we present normalized X-ray line profiles
for various combinations of the parameters $\beta$, $\tau_\ast$, $q$
and $R_o$, and including also the effect of instrumental and/or
macroturbulent broadening as characterized by a Gaussian with a
parameterized width $\sigma$.
We discuss the implications for interpreting observed hot-star X-ray
spectra, with emphasis on signatures for discriminating between
``coronal'' and ``wind-shock'' scenarios.
In particular, we note that in profiles observed so far the
substantial amount of emission longward of line center will be
difficult to reconcile with the expected attenuation by the wind and
stellar core in either a wind-shock or coronal model.
\vspace{0.1in}

\end{abstract}

\section{Introduction}

Over the past two decades, orbiting X-ray observatories from
{\it Einstein} to \asca\/ have shown that both hot and cool stars 
are moderately strong sources of soft X-rays at energies
betweenm 0.1 and 2 keV.
For cooler, late-type stars, the scaling of X-ray luminosity with,
e.g., stellar rotation, suggests a solar-type origin, with surface
magnetic loops confining hot, X-ray-emitting material within nearly
static circumstellar coronae.
For hotter, early-type stars, the X-ray luminosity has no discernible
dependence on rotation, but instead scales nearly linearly with
bolometric luminosity,
$L_{\rm x} \approx 10^{-7} L_{\rm Bol}$ 
\citep{lw80,p81,chs89}, suggesting a different emission mechanism,
perhaps related to the high-density, radiatively driven stellar wind
observed from such stars.

Nonetheless, an ongoing question has remained whether such hot-star
X-ray emission originates from coronal-like processes occuring in
nearly static regions near the stellar surface \citep{co79,w84,ccm97},
or rather in the highly supersonic wind outflow, perhaps from shocks
arising from intrinsic instabilities in the radiative driving
mechanism \citep{lucyw80, l82, ocr88, f97, fpp97, oc99}.
Based on the observed strong winds in hot stars \citep[e.g.,][]{sm76},
a key discriminant has been the apparent general lack of soft X-ray
absorption, which, along with optical and UV spectral constraints,
implies limits to the extent, temperature, and fractional contribution
of the total X-ray output from a base corona
\citep{cos78,nca81,cs83,bl87,m93,c96}.
This limited inferred absorption is thus generally seen as evidence
for emission arising from lower-density regions at large radii, in
general agreement with a wind-shock model.

In the context of extreme ultraviolet spectroscopy, \citet{m91}
presented models of emission line profiles, from both a coronal source
and from an expanding shell, affected by transfer through a stellar
wind.
These authors demonstrated that emission line profiles are a key
discriminant of the source location.
However, due to the small throughput and modest resolution of the {\it
Extreme Ultraviolet Explorer} (\euve) spectrometers, combined with the
extreme opacity of the interstellar medium in this bandpass, emission
lines from only one hot star ($\epsilon$ CMa) were observed with
\euve.
The only high signal-to-noise line seen in this EUV spectrum was the
\ion{He}{2} Ly$\alpha$ line, which was observed to be broadened by
several hundred km s$^{-1}$, lending credence to models that postulate
a wind-origin for the hot plasma in OB stars \citep{c95}.

It is only recently, however, that orbiting X-ray observatories like
\xmm\/ and \chandra\/ have been able to provide the first detections
of spectrally resolved X-ray emission lines from hot, luminous stars
\citep{k01,s01,wc01,cmwmc01}.
Through the broadened profiles of these resolved X-ray emission lines,
these data provide a new, key diagnostic for inferring whether the
X-ray emission originates from a nearly static corona, or from an
expanding stellar wind outflow.  A key goal of the present paper is to
provide a firm basis for interpreting such X-ray emission line
spectra.

Specifically, we describe below the method (\S 2) and results (\S3)
for computations of synthetic line profiles arising from parameterized
forms of X-ray emission in an expanding stellar wind.
In comparison to recent work by \citet{i01}, who derived analytic
forms for X-ray emission profiles for the case of constant expansion
velocity, our analysis takes into account a variation of velocity
$v(r)$ in radius $r$, as parameterized by the so-called `beta-law'
form $v(r)=\vinf (1-\Rstar/r)^\beta$, where $\Rstar$ is the stellar
radius and $\vinf$ is the wind terminal speed.
As in \citet{i01}, the overall level of attenuation is characterized
by an integral optical depth $\tau_\ast$, and the X-ray emission
filling factor is assumed to decline outward as a power law in radius,
$f(r) \sim r^{-q}$.
However, we also allow for this emission to begin at a specified onset
radius $R_o$ that can be set at or above the surface radius $\Rstar$.
We first present scaled emission profiles for various combinations of
the four parameters $\beta$, $\tau_\ast$, $q$, and $R_o/R_\ast$ (\S 3.1).
For selected cases intended to represent roughly the competing models
of a ``base coronal'' emission vs.  instability-generated
``wind-shock'' emission, we next compute synthetic profiles that are
broadened by the estimated instrumental resolution for emission lines
observed by \xmm\/ and \chandra\/ (\S 3.2).
Finally, we discuss (\S 4) how various features of the synthetic
profiles are influenced by properties of the emitting material, and
outline prospects for future work in interpreting hot-star X-ray
spectra.

\section{Computing X-Ray Line Emission from a Spherical Stellar Wind} 
\label{sec:xcomp}

Extended monitoring observations by X-ray missions like \rosat\/
\citep{bs94} have shown that for most early-type stars the X-ray
emission is remarkably constant, with little or no evidence of
rotational modulation, and with overall variations at or below the 1\%
level.
This suggests that the hot, X-ray emitting material must be
well-distributed within the circumstellar volume, whether in a narrow,
near-surface corona, or as numerous ($n>(1/0.01)^2=10^4$), small
elements of shock-heated gas embedded in the much cooler,
UV-line-driven, ambient stellar wind.
Here we model any such complex structure in terms of a spherically
symmetric, smooth distribution of X-ray line emission, with emissivity
$\eta_\lambda (r,\mu)$ at an observer's wavelength $\lambda$ along
direction cosine $\mu$ from a radius $r$.
The resulting X-ray luminosity spectrum $L_\lambda$ is given by
integrals of the emission over direction and radius, attenuated by
continuum absorption within the medium,
\begin{equation}
L_\lambda = 8 \pi^2 \int_{-1}^1 \,d \mu \, 
\int_{\Rstar}^\infty \, dr \, r^2 \eta_\lambda (\mu,r) \Exp^{-\tau [\mu,r]} .
\label{llamdef}
\end{equation}
\noindent

\subsection{Optical Depth Integration}
\label{sec:optdepth}

The absorption optical depth $\tau [\mu,r]$ can be most easily
evaluated by converting to ray coordinates $z \equiv \mu r$ and $p
\equiv r\sqrt{1-\mu^2}$, and then integrating over distance $z$ for
each ray with impact parameter $p$,
\begin{equation}
t [p,z] = \int_z^\infty \kappa \rho [ \rp ] \, d \zp ,
\label{taudef}
\end{equation}
where $\rho [\rp] $ is the mass density at radius $\rp \equiv
\sqrt{p^2+\zp^2}$.  We assume here that the X-ray absorption
cross-section-per-unit-mass, $\kappa$, is a fixed constant that is
independent of wavelength $\lambda$ within the line and radius $r$
within the wind.
For emission at a given $r$ along direction cosine $\mu$, the optical
depth attenuation in eqn.  (\ref{llamdef}) can then be obtained
through the simple substitution,
\begin{equation}
\tau [\mu,r] = t \left[ \sqrt{1-\mu^2} \, r, \, \mu r \right].
\label{t2tau}
\end{equation}

For simplicity, we consider here a steady-state wind with velocity
given by the `beta-law' form $w(r) \equiv v(r)/\vinf =
(1-\Rstar/r)^\beta$.
Taking account also of the occultation by the stellar core of radius
$\Rstar$, we then write
\begin{eqnarray}
t [p,z] 
&=&   \infty   ~~~~~~~~~~~~~~~~~~ ; ~~~ p \le \Rstar ~~\& ~~ z \le \sqrt{\Rstar^2-p^2} 
\nonumber
\\
&=&   
\tau_\ast
\int_z^\infty { \Rstar dz' \over r'^2 (1-\Rstar/r')^\beta }
    ~~~~ ; ~~~ {\rm otherwise} .
\label{taupz}
\end{eqnarray}
Here we have used the mass loss rate, $\Mdot \equiv 4 \pi \rho v r^2$,
to define a characteristic wind optical depth, $\tau_\ast \equiv
\kappa \Mdot /4 \pi \vinf \Rstar$.
Note that in a wind with $\beta=0$ and thus a constant velocity
$v=\vinf$, the radial ($p=0$) optical depth at radius $r$ would be
given simply by $t[0,r] = \tau_\ast \Rstar/r$.
Thus in such a contant-velocity wind, $\tau_\ast$ would be the radial
optical depth at the surface radius $\Rstar$, while $R_1 = \tau_\ast
\Rstar$ would be the radius of unit radial optical depth.

For non-integer $\beta$, the optical depth integral (\ref{taupz}) must
generally be done by numerical quadrature.
But for general integer values, e.g. $\beta=$ 0, 1, 2, or 3, it is
still possible to obtain closed-form analytic expressions for this
optical depth, though the expressions can be quite cumbersome to write
down explicitly (see also \citet{i01}).  
In the present work, these
analytic forms are derived and evaluated using the mathematical
analysis software, {\it Mathematica}.
To provide a specific example, we cite here the analytic result for
the canonical case of $\beta=1$, for which the integral in eqn. 
(\ref{taupz}) becomes
\def\sqp2m1{\sqrt{p^2-1}}
\begin{equation}
{t[p,z] \over \tau_\ast} = 
\left [ {\arctan \left ( { z'/r' \over \sqp2m1 }  \right) +
\arctan \left ( { z \over \sqp2m1 } \right ) 
\over \sqp2m1} \right ]^{z'\rightarrow \infty}_{z'=z} .
\label{taupzb1}
\end{equation}
This expression actually applies under the full general conditions
cited in eqn. (\ref{taupz}), though some care must be taken in its
evaluation for the cases with $p \le 1$.

\begin{figure*}
\epsscale{1.5}
\plotone{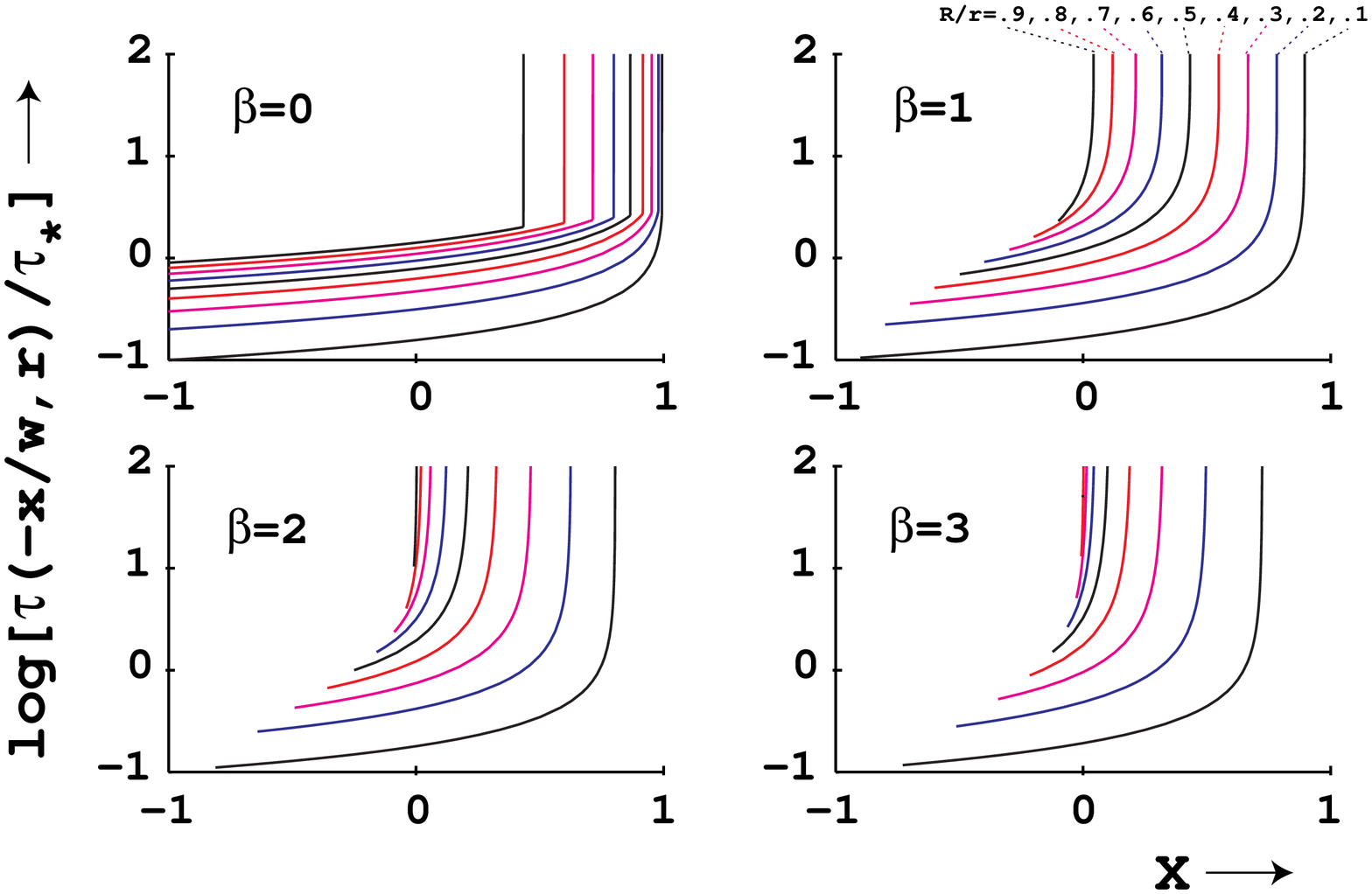}
\caption{
Log of the scaled optical depth $\tau[-x/w(r),r]/\tau_\ast$ vs.  the
scaled wavelength $x$, at selected radii $r$ in models with $\beta=$
0, 1, 2, and 3.
These contours effectively show the degree of wind attenuation as a
function of wavelength for a source at a given radius.
}
\label{fig1}
\end{figure*}

\subsection{Line-Emission Integration}
\label{sec:linem}

Let us next consider the line emissivity, which we take here to be of the form
\begin{equation}
\eta_\lambda (r,\mu) = C \rho^2 (r) f(r) \, \delta \left[ \lambda-\lambda_o(1 - \mu v(r)/c) 
\right] ,
\label{etadef}
\end{equation}
where $C$ is an emission normalization constant, $\lambda_o$ is the
rest wavelength for the line-emission, and $f(r)$ represents a volume
filling-factor for X-ray emission.
Note that the radial dependence of this filling factor may reflect
variations in the temperature distribution of the plasma, as well as
variations in the total emission measure.

The form (\ref{etadef}) simply assumes the emission is an arbitrarily
sharp function centered on the local, Doppler-shifted, comoving-frame
wavelength $\lambda_o (1 - \mu v(r)/c)$, and so ignores all broadening
(e.g., natural, Stark, thermal, turbulent).
For Doppler shifts associated with either thermal or small-scale
``micro-turbulent'' motions, the delta-function should be replaced
with a finite-width emission profile.
But generally, such microscopic motions can be expected to be no
greater than the gas sound speed, which for X-ray emitting plasma of
temperature $T_x \sim 10^7$~K is of order a few times 100~km~${\rm s}^{-1}$.
As such, the associated broadening should be relatively small compared
to that expected from the directed motion of the wind outflow, which
has a much larger characteristic speed $\vinf > 1000$~km~${\rm s}^{-1}$.
Moreover, as discussed in \S 3.2, for larger-scale,
``macro-turbulent'' motions, the effects can be taken into account
{\it a postori} by incorporating them with the general smearing of the
observed line profiles expected from the limited spectral resolution
of the X-ray detector.

For convenience, let us define a scaled wavelength measured from line
center in units of the wind terminal speed $\vinf$, i.e. $ x \equiv
(\lambda/\lambda_o -1) c/\vinf$.
In terms of this scaled wavelength, the integrals (\ref{llamdef}) for
the luminosity spectrum then become
\begin{equation}
L_x = 8 \pi^2 C \int_{-1}^1 \, d\mu \, \int_\Rstar^\infty \, dr \, r^2 \, f(r) \rho^2 (r)
\Exp^{-\tau[\mu,r]} \, \delta (x+\mu w(r)) \, ,
\label{lxdef}
\end{equation}
where we have used the convention $ L_x dx = L_\lambda d\lambda $.
Integrating the Dirac $\delta$-function over direction cosine $\mu$,
we then find that the wavelength-dependent X-ray luminosity can be
evaluated from the single radial integral,
\begin{equation}
L_x = {C \Mdot^2 \over 2 \vinf^2} \int_\Rstar^\infty \, dr \, H \left[ w(r)-|x| \right ] \,
{f(r) \over r^2 
w^3(r)
}
\, \Exp^{-\tau \left [ -x/w(r),r \right]} \, ,
\label{lxrint}
\end{equation}
where $H$ is the Heaviside (a.k.a. unit-step) function, with $H(s)=1$ 
if $s \ge 0$,
and zero otherwise.

Figure \ref{fig1} plots the optical depth $\tau[-x/w(r),r]$ versus
scaled wavelength $x$ at selected radii $r$ in models with various
$\beta$.
For each radius, there is a critical, positive wavelength $x$ where
the optical depth increases steeply, representing the strong
attenuation of the red-wing emission by the dense inner wind and
stellar core.
Such core occultation makes the integrand in eqn.  (\ref{lxq})
effectively zero for all $u$ where $(1-u)^\beta \sqrt{1-u^2} < x$.
For blue-wing wavelengths $x<0$, the termination of the curves
indicates the maximum blue-shift allowed by the velocity at the
relevant radius.

\citet{i01} has recently derived completely analytic results for the
X-ray profiles assuming a power-law emission factor $f(r) \sim r^{-q}$ 
and constant outflow velocity, $\beta=0$ (implying $w(r)=1$).
As noted above, for the somewhat more general case of nonzero, but
{\it integer} values of $\beta$, the optical depth integral
(\ref{taupz}) can still be done analytically;
but in general the radial integral (\ref{lxrint}) must be evaluated by
numerical quadrature.
For this numerical integration, it is convenient (following
\citet{i01}) to define an inverse radius coordinate $u \equiv
\Rstar/r$, yielding
\begin{equation}
L_x = {C \Mdot^2 \over 2 \vinf^2 \Rstar } \int_0^{u_x} \, du 
{f(R_\ast/u) \over (1-u)^{3\beta}}
\, \Exp^{-\tau \left[ -x/(1-u)^\beta, \Rstar/u \right]} \, ,
\label{lxq}
\end{equation}
where $u_x \equiv  1-|x|^{1/\beta}$.\

\section{Results for Normalized Line Profiles from Power Law X-ray Emission} 
\label{sec:xresults}

For any chosen model for the spatial variation of emission $f(r)$, 
eqn. (\ref{lxq}) can be readily evaluated through a single numerical 
integration.
As a specific illustration, we now present results for the
example case of an emission factor with the power-law form 
$f(r) = f_\ast u^q$ for radii $r \equiv \Rstar/u > R_o \equiv \Rstar/u_o$, 
and zero otherwise. 
This cutoff implies a slight redefinition to the upper integral bound in
eqn.(\ref{lxq}),
\begin{equation}
u_x \equiv \min \left [ u_o, 1-|x|^{1/\beta} \right ] \, .
\label{uxdef}
\end{equation}

To focus on the general {\it form} of X-ray line emission,
let us define a normalized emission profile,
\begin{equation}
l_x \equiv {L_x \over \max_x [L_x]} ,
\end{equation}
which simply sets the profile maximum to have a value of unity.
This effectively removes the dependence on the several free parameters
(e.g. $f_\ast$, $\Mdot$, $\vinf$, etc.)  that affect the overall
emission normalization.
The form of the resulting X-ray line-profile $l_x$ versus scaled
wavelength $x$ thus depends on four free parameters, viz.  $\beta$,
$\tau_\ast$, $q$, and $R_o/\Rstar$.
The effects of wind attenuation and other model parameters on the
overall X-ray emission level have been addressed in a broad-band
context by \citet{oc99}, but in focussing here on line profile shapes,
we defer to future work any analysis of the overall emission line
strength.

\begin{figure*}
\epsscale{1.8}
\plotone{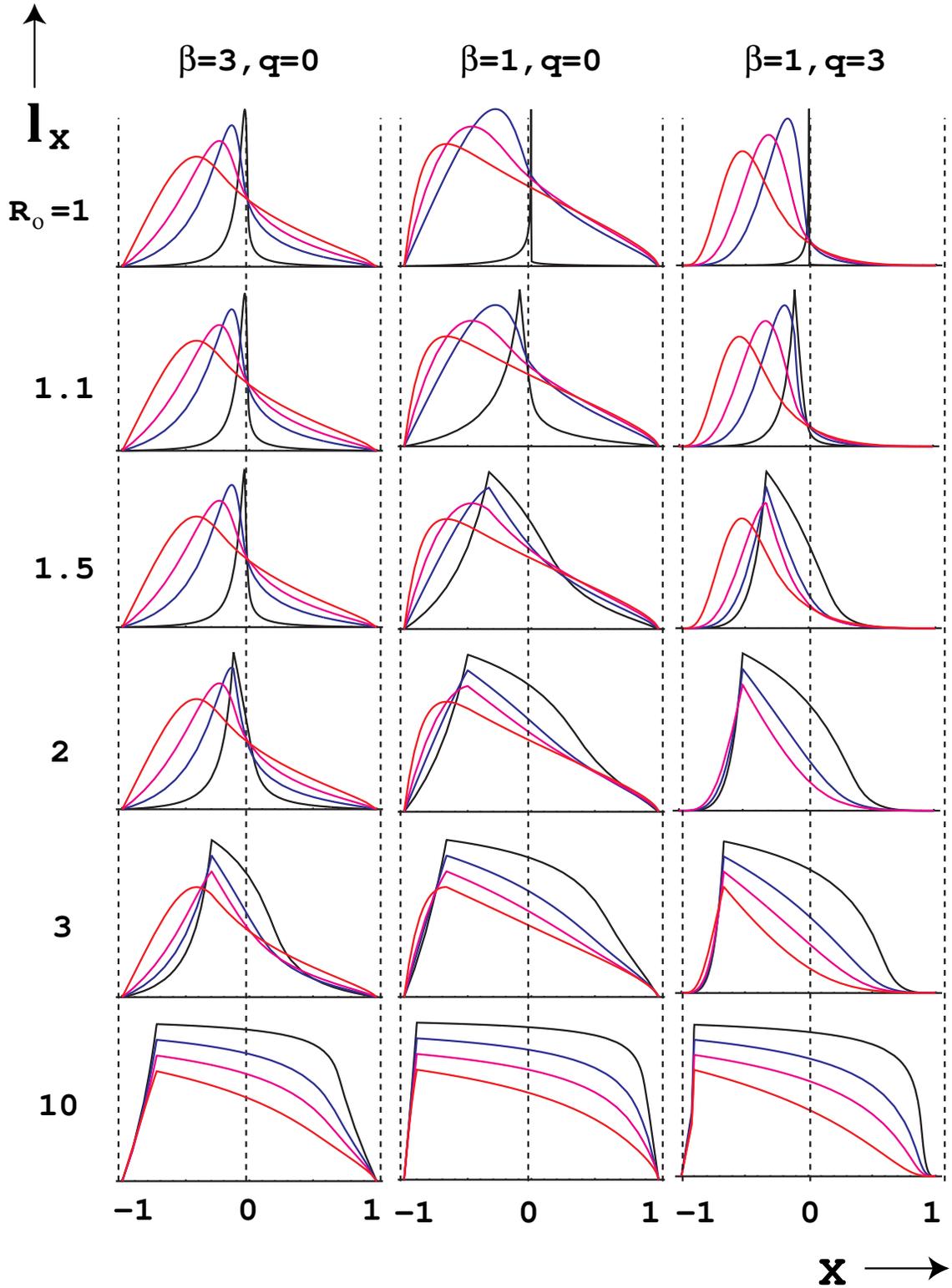}
\caption{
Scaled emission profile $l_x$ vs.  the scaled wavelength $x$, plotted
for various parameter combinations.
In each box, profiles for optical depths $\tau_\ast=$ 1, 3, 5, and 10
are overplotted, with the various curves distinguished by having their
peak values scaled respectively by factors of 1, 0.9, 0.8, and 0.7.
The six rows represent, from the top to bottom, cases with initial
emission radii of $R_o/\Rstar=$ 1, 1.1, 1.5, 2, 3, and 10.
The central of the three columns represents a `standard' case with
$\beta=1$ and $q=0$, while the columns to each side illustrate cases
which differ from this standard by setting $\beta=3$ (left) or $q=3$
(right).}
\label{fig2}
\end{figure*}

\subsection{Synthetic Emission Line Profiles for Various Model Parameters}

\begin{figure*}
\epsscale{1.8}
\plotone{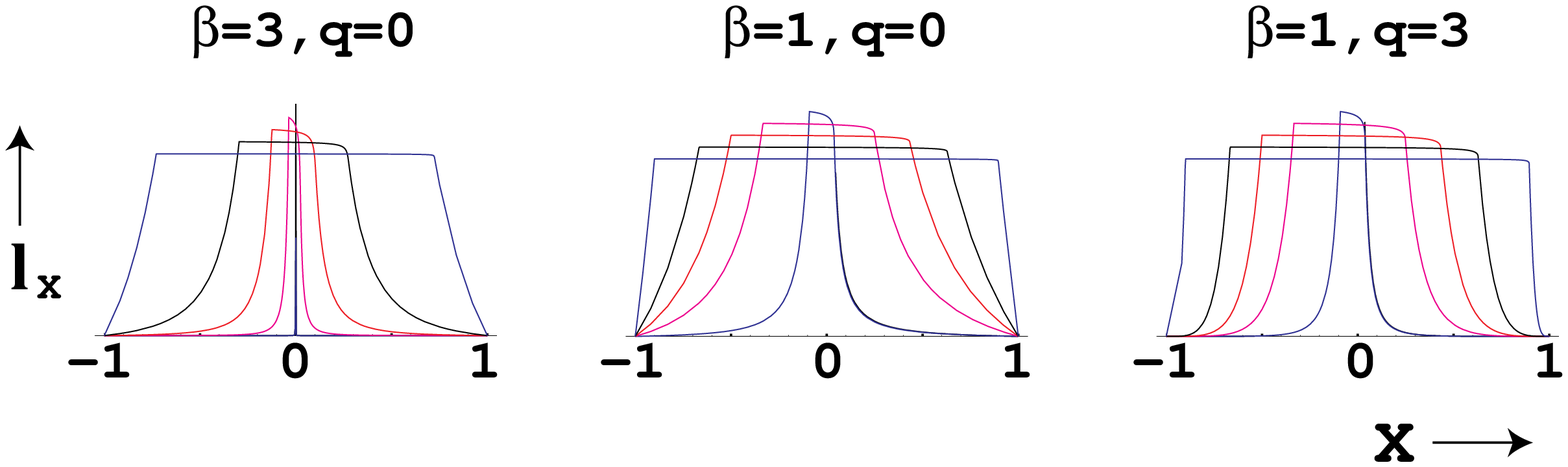}
\caption{
Flat-topped emission profiles for the optically thin case $\tau_\ast=0.01$.
The left, center, and right box correspond to the same wind parameters
as in figure \ref{fig2}, but the five overplots in each box now
represent cases with initial emission radii at $R_o/\Rstar=$ 1.1, 1.5,
2, 3, and 10, with the various curves now distinguished by having
their peak values scaled respectively by factors of 1, 0.95, 0.9,
0.85, and 0.8.
}
\label{fig3}
\end{figure*}

Figure \ref{fig2} presents synthetic emission profiles for various
combinations of these parameters, all plotted vs.  the scaled
wavelength $x$ over the range $-1<x<1$.
For comparison, figure \ref{fig3} shows the flat-topped profiles for
corresponding cases with optically thin emission, $\tau_\ast=0.01$. 
Some notable general properties and trends are as follows:

\begin{itemize}

\item{} The profiles seem most sensitive to $\tau_\ast$ and $R_o$, with $\beta$
and $q$ having a comparatively modest influence.

\item{} The emission peaks are always blueward of line center (i.e., at $x<0$), with 
increasing blueshift for increasing $\tau_\ast$ and/or increasing $R_o$.

\item{} The overall line-width likewise increases with increasing $\tau_\ast$ and/or 
$R_o$. 
The narrowest lines occur when both $\tau_\ast$ and $R_o$ are near
unity.  Large $\tau_\ast$ with low $R_o$ tend to give blue-shifted
``peaked'' forms, while large $R_o$ tend to be blue-tilted
``flat-top'' forms.

\item{} The blue-to-red slope near line center is always negative. 
For low $\tau_\ast$ it varies from very steep for low $R_o$ to nearly
flat for large $R_o$.
For large $\tau_\ast$ it has a more intermediate steepness for all $R_o$.

\end{itemize}

\subsection{Broadened Profiles for Coronal vs. Wind-Shock Scenarios}

For this simple parameterized model, the profiles presented
above represent the expected intrinsic line emission, with the only
broadening associated with the directed, large-scale wind outflow.
In practice, observed line profiles can be further broadened, both by
the limited spectral resolution of the detector, as well as by Doppler
shifting from additional, random motions of the emitting gas.
As already noted in \S 2.2, for small-scale thermal or
``micro-turbulent'' motions, the associated broadening can be expected
to be no larger than the gas sound speed, which for $T_x \sim 10^7$~K
plasma is order a few 100~km~${\rm s}^{-1}$.
Since this is much less than the $v_\infty >1000$~km~${\rm s}^{-1}$ 
characteristic of the directed wind outflow, it should have 
comparitatively minor effect on line-profiles arising from the outer wind.
For the inner wind or a more static ``coronal" models, larger-scale
turbulent broadening, which might in principle have a larger
characeristic speed, can still be accounted for {\it a postori},
together with {\it instrumental broadening} arising from the limited
spectral resolution of the detectors used in the \xmm\/ and \chandra\/
satellites.
For simplicity, we assume here that both macro-turbulent and
instrumental broadening can be characterized by convolving a simple
Gaussian kernel function over the intrinsic line-profile.
As such, a single parameterized width $\sigma$, defined here in units
of the wind terminal speed $\vinf$, represents the combined effects
of both any macroturbulent motion and intrumental broadening, with
$\sigma \equiv \sqrt { \sigma_{turb}^2 + \sigma_{instr}^2 }$.

Let us now consider examples of such broadened line-profiles.
To keep the parameter range tractable, we identify specific parameter
subsets to represent general conditions in the two distinct scenarios
that have often been invoked in modeling hot-star X-ray emission,
namely the ``coronal" \citep[e.g.,][]{w84} vs.  ``wind-shock"
scenarios \citep[e.g.,][]{fpp97,oc99}.
The coronal scenario envisions X-ray emission as coming mostly from a
relatively narrow, nearly static layer near the stellar surface; we
represent this here with the parameter values $q=5$, $\beta=3$, and
$R_o = \Rstar$.
The wind-shock scenario generally envisions X-ray production arising
from instability-generated shocks that form in the acceleration region
$r=1-3 \Rstar$ of the star's line-driven wind, with eventual slow
decay of the shocks in the outer wind;
we represent this here with the parameter values $R_o = 1.5 \Rstar$,
$\beta=1$, and $q=0.5$.
For both scenarios we again consider the effect of various optical
depth parameters, $\tau_\ast$.

\begin{figure*}
\epsscale{1.8}
\plotone{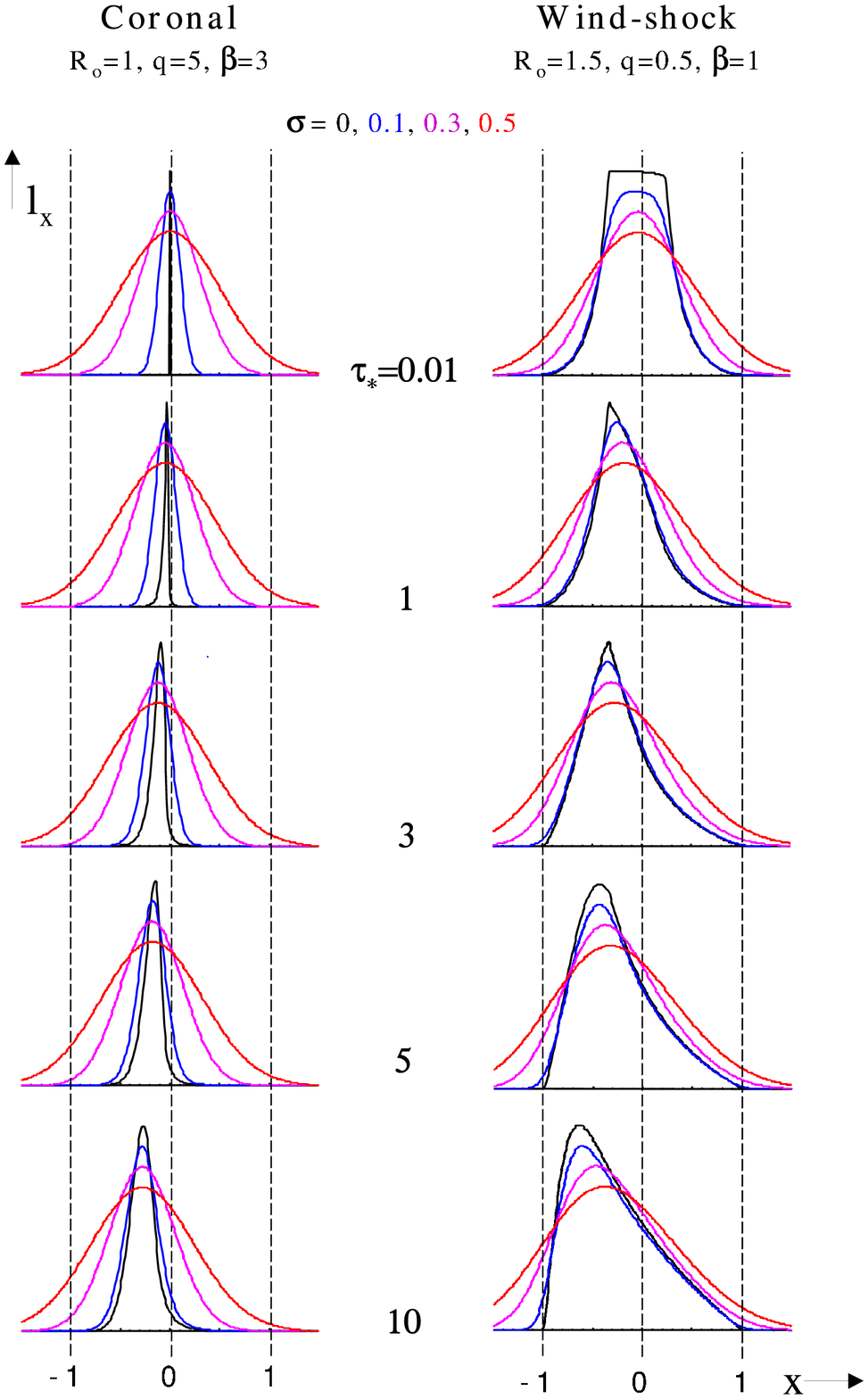}
\caption{
As in figure \ref{fig2}, scaled emission profile $l_x$ vs.  the scaled
wavelength $x$, but now comparing results for a coronal (left) vs. 
wind-shock (right) scenario.
The four overplots now represent profiles with broadening parameters
set to $\sigma=$ 0, 0.1, 0.3 and 0.5, again with the overall emission
scaled respectively by factor of 1, 0.9, 0.8, and 0.7.
The five rows now represent (from the top) cases with optical depth
parameters set respectively to $\tau_\ast=$ 0.01, 1, 3, 5, and 10.  }

\label{fig4}
\end{figure*}

Figure \ref{fig4} compares results for the coronal (left) vs. 
wind-shock (right) scenario.
In characterizing results, we shall refer to profiles with small vs. 
large $\sigma$ as representing good/high vs.  coarse/low resolution.
Here some notable general properties and trends include:

\begin{itemize}

\item{} At the best resolution ($\sigma=0.1$), the coronal models are, as expected, 
notably narrower than the wind models.

\item{} The wind-shock models are generally more blue-shifted
than the coronal models.

\item{} At the coarsest resolution, ($\sigma=0.5$), all profiles have similar
width, but still generally differ in overall blue-shift.

\item{} For the optically thin case, the emission is nearly symmetric about 
line center. For other cases, the emission peaks become more blue-shifted 
with increasing optical  depth $\tau_\ast$.  
(The  coronal models with larger $\tau_\ast$ may not be relevant, however, 
since wind absorption would greatly reduce the overall strength of any 
observable emission.)

\end{itemize}

The typical spectral resolution for \chandra\/ is around 
$\lambda/\Delta \lambda \sim 1000-1500$.
For characteristic
wind terminal speed of $\vinf \sim 2000-3000$, this implies a 
typical broadening parameter of $\sigma \sim 0.1$.
For \xmm\/, the spectral broadening is comparable, though slightly larger.
Clearly, at such high resolutions, it should be possible to place strong
constraints on the site of X-ray emission and the degree of wind
attenuation, as well as on the characteristic speeds of directed vs. 
turbulent motions of the emitting plasma.

\section{Discussion}

The parameterized emission models here provide a good basis for
interpreting X-ray line-profile observations made possible by the new
generation of orbiting X-ray telescopes.
While detailed, quantitative analyses must await the general
availability of specific datasets, it is already clear from the few
hot-star X-ray spectra published thus far that there are some
noteworthy challenges for modeling the associated X-ray emission.

For example, temperature-sensitive line-ratio diagnostics provide
constraints on the magnitude of shocks in dynamical models.
Generally instability generated wind-shock models tend to be dominated
by moderate shock-strengths $\Delta v \sim 100-400$~km~${\rm s}^{-1}$ 
that can only heat material up to a maximum of $\sim 10^7$~K \citep{f97}.
In those stars with much higher inferred temperatures, for example
$\theta^1$ Ori C with $T > 5 \times 10^7$ K \citep{s01}, there is a need
to produce stronger shocks.
One possibilty is for wind outflow to be channeled by magnetic 
loops \citep{bm97}.
In this case, wind material accelerated up opposite footpoints of a closed
loop can be forced to collide near the loop top, in principle yielding
shocks with velocity jumps of order $\Delta v \sim 1000$~km~${\rm s}^{-1}$, 
a substantial fraction of the wind terminal speed.

Another key diagnostic stems from the forbidden-to-intercombination (f/i) 
ratios of helium-like emission lines, which can provide a strong constraint on 
the location of the X-ray emission.
In the case of the solar corona, such ratios are nominally interpreted
as density diagnostics \citep{gj69}.
But because the strong UV radiation of hot-stars can generally
dominate over electron collisions in the destruction of forbidden line
upper levels, in such stars the f/i provides instead a constraint on
the strength of the UV intensity \citep{bdt72, ms78}.
With appropriate atmospheric modeling of the relevant part of UV
continuum, any inferred dilution of the radiation thus provides a
diagnostic of the proximity of the X-ray emission to the stellar surface.
For \xmm\/ observations of $\zeta$~Pup \citep{k01}, initial
interpretations suggest formation well away from the stellar surface
($r>2 \Rstar$), consistent with a wind-shock origin, with moreover a
general trend toward larger radii for lower-energy lines, as would be
expected from the larger optical depths and thus larger
unit-optical-depth radii for such lines.
For Chandra observations of $\zeta$~Ori \cite{wc01}, the inferred
formation radii are also generally away from the star ($r \approx 2-10
\Rstar$), but with one stage (\ion{Si}{13}) suggesting a formation very
close to the surface ($r < 1.1 \Rstar$).
However, in this latter case, the relevant UV flux for destruction
of the forbidden upper levels lies at energies above the
Lyman-edge, and so any error in atmospheric modeling  of the Lyman jump 
could still alter this inferred formation radius.
Chandra observations of $\zeta$~Pup \citep{cmwmc01} show
similar results, with most inferred formation radii in the range
$r \approx 1.5-10 \Rstar$, with trends again consistent with inferred 
unit-optical-depth radii, but again with one stage (in this case 
\ion{S}{15}) implying formation quite close to the surface ($r < 1.25 \Rstar$).
Of course, such fir constraints on source location must eventually be 
considered in conjunction with additional constraints, such as those 
provided by analysis of line profiles. 

Indeed, in the context of this paper, the most notable new diagnostic provided
by recent X-ray observations lies in the form of the line profiles.
For some initial results, these have some quite puzzling characteristics.
For example, in the \xmm\/ spectrum of $\zeta$ Pup \citep{k01}, the
lines all exhibit a similar, significant broadening, but with
half-widths $\sim 1000-1500$~km s$^{-1}$ that are still somewhat lower
than the inferred wind terminal speed $\vinf \gtwig 2000$~km s$^{-1}$.
The typical profiles have a rounded, Gaussian form that is quite
distinct from, e.g., the flat-topped profiles expected for optically
thin emission from a nearly constant velocity expansion.
Moreover, the profiles show only modest asymmetry, with the peak intensity only
slightly blue-shifted, by $\sim 0-400$~km s$^{-1}$ relative to
line-center, and the red-wing only slightly flatter and more extended
than the blue-wing.
Recent somewhat higher spectral resolution observations by Chandra do
show the greater shift and asymmetry expected from optically thick
wind emission for $\zeta$ Pup \citep{cmwmc01}.
But in Chandra observations of the O9.7 Ib star $\zeta$ Ori, the
profiles are again very broad (half-widths of $\sim$~1000~km~${\rm s}^{-1}$), 
but nearly symmetric \citep{wc01}.

Such profiles with substantial broadening, but little asymmetry and
blueshift are difficult to reconcile with a simple wind-outflow model
that includes attenuation by the wind and stellar core, both of which
should preferentially attenuate X-rays from the back-side hemisphere,
and thus in a wind or a corona with some outflow, preferentially
reduce the {\it red}-side emission.
In fact, from detailed modeling constrained by \rosat\/ observations,
\citet{h93} have argued that that the wind of \zpup\/ should be
optically thick to 1 keV X-rays at 10 \rstar, with the radius of
optical depth unity being even greater for softer X-rays (20 \rstar\/
at 0.6 keV; and still 2 \rstar\/ at 2 keV).

Even for a base corona with some overall outflow velocity and {\it no}
overlying wind attenuation, occultation by the stellar core should
still produce line profiles that are skewed toward the blue.
A coronal model with little or no net outflow could explain the
observed profile symmetry, but matching the observed line-widths would
then require the coronal material to have extremely large
``turbulent'' motions.  
For example, line widths in the Chandra
observations of $\zeta$~Ori \citep{wc01} require velocity dispersions
of order 1000~km~${\rm s}^{-1}$, and some of the lines observed by \chandra\/ in
$\theta^1$ Ori C \citep{s01} are even broader than this, yet
apparently still nearly symmetric.
If the lack of red-wing attenuation is interpreted to mean that red-shifted
emission does not arise from wind outflow from the backside hemisphere,
then this suggest that material {\it in front of the star}
must be flowing {\it downward} at velocities of 1000~km~${\rm s}^{-1}$.  
Since this is greater than the surface escape speed, it cannot be the
result of gravitational infall, but instead would imply that
some mechanism, e.g., perhaps magnetic reconnection \citep{wc01}, must
actively accelerate roughly equally strong upflow and downflow with
speeds of 1000~km~${\rm s}^{-1}$!
Moreover, such random motions should inevitably lead to high-speed
gas collisions and their associated shock heating.
For shocks with velocity jumps of order 1000~km~${\rm s}^{-1}$ the associated
post-shock temperature should be of order $10^8$~K, implying generally 
harder X-rays and higher ionization stages than are typically observed.
In this context, these broad but symmetric profiles would indeed seem to
represent a strong challenge for theoretical models.

On the other hand, our synthesized line profiles from {\it wind}
models do seem capable of producing relatively broad, yet symmetric
and unshifted profiles, but only in the nearly complete absence of
wind attenuation, and with emission originating far from the
occultation of the stellar core.
To test the viabilty of this alternative, observed line-profiles should
be interpreted in conjunction with other diagnostics, like the f/i 
ratio, that can constrain the radii of X-ray emission, and with an
accurate account of wind absorption, as computed through opacity
calculations for the observationally inferred wind mass loss rate.

Any wind shock model with significant wind attenuation is expected to
be strongly blueshifted and quite asymmetric (see figure \ref{fig4}).
Of all the line profiles discussed in the first group of \xmm\/ and
\chandra\/ papers, the observations of $\zeta$~Pup provide the best
evidence for the blue-shifts and asymmetry expected for optically
thick wind emission \citep{k01,cmwmc01}.
For example, figure 3 of Cassinelli et al. (2001) shows 5 out of 6 lines with 
blueward shifts and asymmetry similar to profiles calculated here for
wind-shock models with substantial optical depth, e.g. $\tau_\ast \sim 5$.
At least for this canonical O-type star, these profiles seems to provide strong 
evidence of a wind origin for most of the observed X-rays.

In summary, systematic fitting of the line-profiles will be necessary
to place more quantitative constraints on the spatial distribution of
hot plasma and the degree of wind attenuation in the observed O stars;
but some initial qualitative inferences are already possible.
For some cases, the implied minimal wind attenuation combined with some 
significant bulk motion are puzzling, both in the context of coronal X-ray
emission and in the context of the strong radiation-driven winds that
are known to exist on these stars from UV, optical, IR, and radio
observations.
One possible solution to the wind attenuation problem might be found
in clumping.  A ``porous'' wind might allow many of the X-rays to
stream through unattenuated, yet also lead to significant UV
absorption if the interclump wind medium still has sufficient column
density to produce optically thick UV lines.
(The UV line transition probabilities are significantly larger
than the bound-free X-ray cross sections.) 
An alternative is that the mass-loss rates of O stars may have been
overestimated.  
It does not seem likely, however, that the winds are dense
and smooth but transparent to X-rays, as that would require levels of
ionization in the wind that would not allow for efficient
radiation-driving.
And, if the winds are indeed optically thin, then to discriminate
between coronal and wind models will require a careful analysis of the
shifts and asymmetries observed in the X-ray line profiles.

Finally, although we developed our analysis with hot stars in mind, it
could be extended to other X-ray sources in which lines are emitted within a
nearly symmetric expanding medium that may also have substantial
continuum opacity.
This could include nova winds, planetary nebulae, active galactic
nuclei, and X-ray binaries.
  
\acknowledgements
We thank R. Ignace for providing a preprint of his paper on analytic
X-ray profiles and also for his detailed comments on this manuscript.
We also thank K. Gayley and A. Ud-Doula for helpful comments on our
analysis, and J. Cassinelli and J. MacFarlane for many useful insights
regarding the X-ray spectra of O stars.
Finally, we thank the referee, M. Corcoran, for helpful comments and
suggestions on the submitted manuscript.
This research was supported in part by NASA grant NAG5-3530 to the
Bartol Research Institute at the University of Delaware.



\begin{thebibliography}{}

\bibitem[Baade \& Lucy(1987)]{bl87} Baade, D., \& Lucy, L. 1987, \aap, 178, 213

\bibitem[Babel \& Montmerle (1997)]{bm97} Babel, J. \& Montmerle, T. 1997, \apjl, 485, L29

\bibitem[Bergh\"ofer \& Schmitt(1994)]{bs94} Bergh\"ofer, T. W., \& Schmitt, J. H. M. M.
1994, \apss, 221, 309

\bibitem[Blumenthal, Drake, \& Tucker(1972)]{bdt72} Blumenthal, G. R., Drake, G. W. F., \&
Tucker, W. H. 1972, \apj, 172, 205

\bibitem[Cassinelli et al.(1995)]{c95} Cassinelli, J. P., Cohen, D. H., Macfarlane, J. J., 
Drew, J. E., Lynas-Gray, A. E., Hoare, M. G., Vallerga, J. V., Welsh, B. Y., 
Vedder, P. W., Hubeny, I., \& Lanz, T. 1995, \apj, 438, 932

\bibitem[Cassinelli et al.(2001)]{cmwmc01} Cassinelli, J. P., Miller, N. A., Waldron, W. L., 
MacFarlane, J. J., \& Cohen, D. H. 2001, \apj, in press

\bibitem[Cassinelli \& Olson(1979)]{co79} Cassinelli, J. P., \& Olson, G. L. 1979, \apj, 229, 304

\bibitem[Cassinelli, Olson, \& Stalio(1978)]{cos78} Cassinelli, J. P., Olson, G. L., \& Stalio, R. 
1978, \apj, 220

\bibitem[Cassinelli \& Swank(1983)]{cs83} Cassinelli, J. P., \& Swank, J. H. 1983, \apj, 271, 681

\bibitem[Chlebowski, Harnden, \& Sciortino(1989)]{chs89} Chlebowski, T., 
Harnden, F. R., Jr., \& Sciortino, S. 1989, \apj, 341, 427

\bibitem[Cohen et al.(1996)]{c96} Cohen, D. H., Cooper, R. G., 
MacFarlane, J. J., Owocki, S. P., Cassinelli, J. P., \& Wang, P. 1996, \apj, 460, 506

\bibitem[Cohen, Cassinelli, \& MacFarlane(1997)]{ccm97} Cohen, D. H., Cassinelli, J. P., 
\& MacFarlane, J. J. 1997, \apj, 487, 867

\bibitem[Feldmeier et al.(1997)]{f97} Feldmeier, A., Kudritzki, R.-P., Palsa, R., 
Pauldrach, A. W. A., \& Puls, J. 1997, \aap, 320, 899

\bibitem[Feldmeier, Puls, \& Pauldrach(1997)]{fpp97} Feldmeier, A., 
Puls, J., \& Pauldrach, A. W. A. 1997, \aap, 322, 878

\bibitem[Gabriel \& Jordan(1969)]{gj69} Gabriel, A. H., \& Jordan, C. 1969, \mnras, 145, 241

\bibitem[Hillier et al.(1993)]{h93} Hillier, D. J., Kudritzki, R. P., Pauldrach, A. W., Baade, D., 
Cassinelli, J. P., Puls, J., \& Schmitt, J. H. M. M. 1993, \aap, 276, 117

\bibitem[Ignace(2001)]{i01} Ignace, R. 2001, \apj, in press

\bibitem[Kahn et al.(2001)]{k01} Kahn, S. M., Leutenegger, M. A., Cotam, J., Rauw, G., 
Vreux, J.-M., den Boggende, A. J. F., Mewe, R., \& G\"udel, M.  2001, \aap, in press

\bibitem[Long \& White(1980)]{lw80} Long, K. S., \& White, R. L. 1980, \apj, 239, L65

\bibitem[Lucy(1982)]{l82} Lucy, L. 1982, \apj, 255, 286

\bibitem[Lucy \& White(1980)]{lucyw80} Lucy, L. B., \& White, R. L. 1980, \apj, 241, 300

\bibitem[MacFarlane et al.(1991)]{m91} Macfarlane, J. J., Cassinelli, J. P., Welsh, B. Y., 
Vedder, P. W., Vallerga, J. V., \& Waldron, W. L. 1991, \apj, 380, 564

\bibitem[MacFarlane et al.(1993)]{m93} MacFarlane, J. J., Waldron, W. L., Corcoran, M. F., Wolff, 
M. J., Wang, P., \& Cassinelli, J. P. 1993, \apj, 419, 813

\bibitem[Mewe \& Schrijver(1978)]{ms78} Mewe R., \& Schrijver, J. 1978, \aap, 65, 99

\bibitem[Nordsieck, Cassinelli, \& Anderson(1981)]{nca81} Nordsieck, K. H., Cassinelli, J. P., 
\& Anderson, C. M. 1981, \apj, 248, 678

\bibitem[Owocki, Castor, \& Rybicki(1988)]{ocr88} Owocki, S. P., Castor, J. I., \& Rybicki, G. B. 
1988, \apj, 335, 914

\bibitem[Owocki \& Cohen(1999)]{oc99} Owocki, S. P., \& Cohen, D. H. 1999, \apj, 520, 833

\bibitem[Pallavicini et  al.(1981)]{p81} Pallavicini, R., Golub, L., Rosner, R., 
Vaiana, G. S., Ayres, T., \& Linsky, 

J. L. 1981, \apj, 248, 279
\bibitem[Schulz et al.(2001)]{s01} Schulz, N. S., Canizares, C. R., 
Huenemoerder, D., \& Lee, J. C. 2001, \apjl, in press

\bibitem[Snow \& Morton(1976)]{sm76} Snow, T. P., Jr., \& Morton, D. C. 1976, \apjs, 32, 429

\bibitem[Waldron(1984)]{w84} Waldron, W. L. 1984, \apj, 282, 256

\bibitem[Waldron \& Cassinelli(2001)]{wc01} Waldron, W. L., \& Cassinelli, J. P. 2001, \apjl, 
in press

\end{thebibliography}
\end{document}